\documentclass{article}
\newcommand{\beq}{\begin{equation}}
\newcommand{\eeq}{\end{equation}}
\newcommand{\beqa}{\begin{eqnarray}}
\newcommand{\eeqa}{\end{eqnarray}}
\newcommand{\beqar}{\begin{eqnarray*}}
\newcommand{\eeqar}{\end{eqnarray*}}

\newcommand{\labell}[1]{\label{#1}} 
\newcommand{\reef}[1]{(\ref{#1})}

\newcommand{\eg}{{\it e.g.,}\ }
\newcommand{\ie}{{\it i.e.,}\ }

\newcommand{\norm}[1]{\raise.3ex\hbox{:}#1\raise.3ex\hbox{:}}

\newcommand{\tr}{{\rm tr}}

\newcommand{\D}{\Delta}

\newcommand\cL{{\cal L}}

\newcommand\cR{{\cal R}}

\newcommand\tC{{\widetilde C}}

\newcommand{\half}{\frac{1}{2}}

\newcommand{\bi}{\begin{itemize}}
\newcommand{\ei}{\end{itemize}}
\newcommand{\ben}{\begin{enumerate}}
\newcommand{\een}{\end{enumerate}}

\newcommand{\const}{\mbox{const}}
\newcommand{\bmtx}{\left[ \begin{array}{cc}}
\newcommand{\emtx}{\end{array} \right]}
\newcommand{\bvec}{\left[ \begin{array}{c}}
\newcommand{\evec}{\end{array} \right]}

\newcommand{\mysection}[1]{\section{#1} \setcounter{equation}{0}}

\newcommand{\RR}{{\mathcal R}}
\newcommand{\QQ}{{\mathcal Q}}
\renewcommand{\AA}{{\mathcal A}}
\newcommand{\TT}{{\mathcal T}}
\newcommand{\II}{{\mathcal I}}

\newcommand{\bfig}{\begin{figure}} 
\newcommand{\efig}{\end{figure}}
\newcommand{\nn}{\nonumber}
\newcommand{\ba}{\begin{array}}
\newcommand{\ea}{\end{array}}

\newcommand{\bs}{\begin{slide}}
\newcommand{\es}{\end{slide}}

\renewcommand{\D}{n}  

\renewcommand{\newline}{\\}
\renewcommand{\tC}{\widetilde{C}}

\newtheorem{theorem}{Theorem}

\begin{document}

\title{A note on spherically symmetric naked singularities in general
dimension}
\author{Peter Langfelder and Robert B. Mann  \and  %EndAName
\textit{Dept. Of Physics, University of Waterloo, Waterloo,
ON N2L 3G1 Canada }  \and  \texttt{plangfel@uwaterloo.ca,
mann@avatar.uwaterloo.ca} }
\date{\today }
\maketitle

\begin{abstract}
We discuss generalizations of the recent theorem by Dafermos
(hep-th/0403033) forbidding a certain class of naked singularities in the
spherical collapse of a scalar field. Employing techniques similar to the
ones Dafermos used, we consider extending the theorem (1) to higher
dimensions, (2) by including more general matter represented by a
stress--energy tensor satisfying certain assumptions, and (3) by replacing
the spherical geometry by a toroidal or higher genus (locally hyperbolic)
one. We show that the extension to higher dimensions and a more general
topology is straightforward; on the other hand, replacing the scalar field
by a more general matter content forces us to shrink the class of naked
singularities we are able to exclude. We then show that the most common
matter theories (scalar field interacting with a non-abelian gauge field and
a perfect fluid satisfying certain conditions) obey the assumptions of our
weaker theorem, and we end by commenting on the applicability of our results
to the five-dimensional AdS scenarii considered recently in the literature. 
%, namely we can only exclude non-integrable singularities that lie in the regular region.
\end{abstract}

\mysection{Introduction}

The issue of cosmic censorship has received renewed attention lately,
sparked by claims in the literature~\cite{HHM1, HHM2} that there are
physically well-motivated theories in which \emph{open} sets of initial
conditions lead to the formation of a naked singularity. Several groups~\cite%
{kraus, shenker, garfinkle, frolov} have attempted to find explicit examples
of such configurations, but were not successful; further, Dafermos~\cite%
{dafermos} has shown that in $\D=4$ dimensions, naked singularities of the
type predicted by~\cite{HHM1} in fact cannot arise.

In this paper we employ techniques similar to those used by Dafermos to
study whether a similar statement could be made for gravity coupled to a
more general matter content in an arbitrary dimension. We further generalize
by including singularities of toroidal and hyperbolic topology. Our matter
content will be represented by a stress tensor $T_{\mu\nu}$ satisfying
certain conditions inspired by the scalar system considered by Dafermos. We
will see that while a higher-dimensional generalization including more
general topologies of the singularity is straightforward, a theorem similar
to the one Dafermos formulated cannot be proven without a detailed knowledge
of the dynamics of the matter content. Nevertheless, one can make some
progress and exclude certain types of singularities even with a general $%
T_{\mu\nu}$.

Before proceeding with the details of this generalization, it should be
pointed out that our results do not directly apply to the 5-dimensional
supergravity scenarios of~\cite{HHM2} and~\cite{HH2}, since their work uses
a scalar potential unbounded from below that violates the assumptions we
will make. It would be interesting to see under what circumstances (if any)
the theorem proved here could be generalized for such unbounded potentials.
We comment on this issue briefly in Sec.~\ref{sec:comments}.

This paper is structured as follows. In Section~\ref{sec:EOMs} we describe
the system we will study, discuss the assumptions on which our results
depend and introduce some useful notation. In Section~\ref{sec:Theorem} we
review the formulation of Dafermos' theorem and attempt to generalize its
proof to our more general system. While we are able to make some progress,
at a certain point we are forced to either make additional restrictions on
the kind of singularities that we can exclude or we must restrict the matter
content such that we can make progress along the lines of the original proof
by Dafermos. We discuss the former, along with some examples of matter
content that satisfies our assumptions, in Section~\ref%
{sec:GeneralSingularities}, while the latter is analyzed in Section~\ref%
{sec:Scalar}. We close in Sec.~\ref{sec:comments} with comments on possible
extensions and applications to the case of unbounded scalar potential.

\mysection{Setup and assumptions}

\label{sec:EOMs}

We work in $\D$ dimensions ($\D\geq 4$) and consider a system of
Einstein--Hilbert gravity coupled to  matter with stress tensor $T_{\mu \nu }
$. The action is \beq S = \int d^{\D} x \thinspace\ \sqrt{g} \left[ \half %
R+\cL_{matter}\right]  \thinspace . \labell{action} \eeq For now we leave
the matter Lagrangian general. Our notation is standard with $R$ denoting
the scalar curvature of the metric $g$. The gravitational equations of
motion are \beq R_{\mu \nu }- \half g_{\mu \nu }R = 2T_{\mu \nu }, %
\labell{GrEOM} \eeq with $T_{\mu \nu }$ denoting the matter stress tensor.

Dafermos~\cite{dafermos} gave a thorough discussion of the structure of the
manifold described by the metric $g$. Our assumptions are the same except
for allowing the ``transverse part'' of the manifold to be somewhat more
general. We will work with light-cone coordinates $u,v$ plus a transverse $(%
\D-2)$-dimensional space. The metric of the latter will be denoted by $%
\gamma^{(\D-2)}$, and we will take it to be Einsteinian ($R_{ij}^{(\D-2)}=k(\D%
-3)\gamma _{ij}^{(\D-2)})$ with the constant $k$ taking one of the values%
\footnote{%
One could take $k$ to be any other real constant as well; the relevant
distinction is in the sign of $k$.} $1,0,$ or $-1$. The full space-time
metric will thus have the form  
\beq
ds^{2}=-\Omega ^{2}dudv+r^{2}\gamma^{(\D-2)}\labell{Metric}
\eeq
where $\Omega $ and $r$ are functions of $u,v$ only.

The basic assumptions employed by Dafermos (and us) are (1) that the
manifold be evolutionary (roughly speaking, this means that its time
evolution is fully determined by suitable initial conditions) and (2) that
there is enough symmetry in the transverse coordinates to effectively reduce
the evolution to a 2-dimensional problem represented by a region $\QQ$
parametrized by the coordinates $u, v$.

\subsection{Matter Stress--Energy Tensor}

The matter stress tensor will be assumed to satisfy the following
inequalities, written in the coordinate frame of~\reef{Metric}: \beqa 
T_{uu} & \geq & 0 \nn \newline
T_{vv} & \geq & 0 \labell{ConditionsOnT} \newline
T_{uv} & \geq & -|2g_{uv}| C\, , \nn \eeqa where $C$ is a
constant that we will take to be positive. We will further assume that all
components $T_{\mu\nu}$ are continuous. These conditions are a direct
generalization of the ones Dafermos~\cite{dafermos} imposed on the scalar
potential in a system of gravity coupled to a scalar field. Additionally we
assume that the matter stress tensor respects the symmetry of the transverse
coordinates, meaning that the internal coordinates of the stress tensor must
be proportional to the internal metric, with the proportionality factor
independent of the internal coordinates (though it can be a function of $u,v$%
). This is a natural assumption in the light of the fact that we are
assuming a metric that is (maximally) symmetric in the internal coordinates;
indeed, it is hard to imagine a maximally symmetric metric being sourced by
a matter stress tensor that would not have this symmetry.

As an aside, we note that the conditions~\reef{ConditionsOnT} imply a
modified form of the weak energy condition: we have  
\beq
T_{\mu \nu }\xi ^{\mu }\xi ^{\nu }\geq 2C\xi ^{2}\labell{WeakECond}
\eeq%
for any non-spacelike $\xi $ (\ie$\xi ^{2}\leq 0$) whose components lie only
in the $u,v$ directions\footnote{%
We remind the reader that the conventional weak energy condition has the
form 
\[
T_{\mu \nu }\xi ^{\mu }\xi ^{\nu }\geq 0 \nn
\]
for any non-spacelike vector $\xi $. Details can be found \eg in~\cite%
{HawkingBook}.}. However it is not possible to derive the conditions~%
\reef{ConditionsOnT} from~\reef{WeakECond}. Likewise, the conditions~%
\reef{ConditionsOnT} are not strong enough to derive an analogue of the
dominant and/or strong energy conditions.

\subsection{Geometry of $\QQ$ and the Extension Criterion}

We again refer the reader to Ref.~\cite{dafermos} for an in-depth discussion
of the geometry of $\QQ$ and the extension criterion characterizing the
boundary of $\QQ$; for our purposes it will be enough to summarize the main
points. The metric on $\QQ$ is $-\Omega^2 du dv$, with $u,v$ being global
coordinates. The region $\QQ$ is assumed to have a boundary consisting of a
spacelike curve $S$ and a timelike curve $\Gamma$, whose intersection is a
single point. The curve $\Gamma$ consists of all points with $r=0$, and is
called the \emph{centre}. One also defines the \emph{infinity} $\II$ and
ingoing ($u=\const$) and outgoing ($v=\const$) null curves.

Denoting \beqa \nu &\equiv& r_{,u} \labell{DefnOfNu} \newline
\lambda &\equiv& r_{,v} \labell{DefnOfLambda} 
\eeqa we assume that we have $\nu <0$ along the spacelike boundary $S$ of $%
\QQ$ and at infinity\footnote{ The boundary condition at infinity is necessary 
\eg in asymptotically AdS
spacetimes where the spacelike boundary does not constitute a Cauchy surface.} $\II$.
From the $uu$ component of the equations of motion, \beq \left(
\Omega^{-2} \nu \right),_u = -\frac{2}{\D-2} r T_{uu} \eeq and the
assumption $T_{uu}>0$ we then deduce \beq  \nu<0 \mbox{\ \ \ everywhere in \
}\QQ\, . \eeq The regular region $\RR$, the trapped region $\TT$ and the
marginally trapped region $\AA$, are defined, respectively, as \beqa \RR &=&
\left\{ q\in \QQ: \lambda(q)>0 \right\} \, , \nn \newline
\TT &=& \left\{ q\in \QQ: \lambda(q)<0 \right\} \, , \labell{RegionDefs} \newline
\AA &=& \left\{ q\in \QQ: \lambda(q)=0 \right\} \, . \nn %
\eeqa  To give a short summary of the extension criterion as formulated by
Dafermos, we must introduce some more terminology. For any set $U$ in the $%
(u,v)$ plane, we denote by $\overline U$ the closure of $U$ in the topology
of the plane. The symbols $J^+$ and $J^-$ refer to the causal future and
past (\ie regions reachable by non-spacelike curves), respectively, while $%
D^+$ denotes the domain of dependence and $I^+, I^-$ to the chronological
future and past (\ie regions reachable by timelike curves\footnote{%
Precise definitions can be found \eg in Hawking and Ellis~\cite{HawkingBook}.%
}).

For a point $p \in \overline{\QQ}$ we will call the indecomposable past (IP)
subset $J^-(p) \cap \QQ$ \emph{eventually compactly generated} iff there
exists a compact subset $X \subset \QQ$ such that \beq  J^-(p) \subset
D^+(X) \cap J^-(X) \, . \labell{DefnOfCompactlyGenerated} \eeq A point $p\in
\overline\QQ \setminus \QQ$ will be called a \emph{first singularity} iff $%
J^-(p) \cap \QQ$ is eventually compactly generated and if any eventually
compactly generated indecomposable proper subset of $J^-(p) \cap \QQ$ is of
the form $J^-(q)$ for some $q \in \QQ$.

This definition means that if $p=(u_{s},v_{s})\notin \overline{\Gamma }$ is
a first singularity, there exists an $\epsilon >0$ such that, denoting $%
u_{\epsilon }\equiv u_{s}-\epsilon ,v_{\epsilon }\equiv v_{s}-\epsilon $,
the compact set 
\beq
X=\{u_{\epsilon }\}\times \lbrack v_{\epsilon },v_{s}]\cup \lbrack
u_{\epsilon },u_{s}]\times \{v_{\epsilon }\}
\eeq
satisfies $X\subset \QQ\setminus \Gamma $, and  
\beq
\lbrack u_{\epsilon },u_{s}]\times \lbrack v_{\epsilon
},v_{s}]=D^{+}(X)=J^{-}(p)\cap D^{+}(X)
\eeq%
\thinspace\ while  
\beq
D^{+}(X)\cap \QQ=D^{+}(X)\setminus \{p\}
\eeq
For $Y\subset \QQ\setminus \Gamma $ we define a ``norm''  
\beq
N(Y)=\sup \{|\Omega |_{1},|\Omega ^{-1}|_{0},|r|_{2},|r^{-1}|_{1}\}\,,
\eeq
where, for a function $f$ defined on $\QQ$, $|f|_{k}$ denotes the
restriction of the $C^{k}$ norm to $Y$. For any \emph{compact} $Y\subset \QQ$
disjoint with the centre $\Gamma $ the norm $N(Y)$ is finite.

The extension criterion can then be formulated as the following property,
which we will take as an assumption\footnote{%
Dafermos~\cite{dafermos} derives the property from standard results of the
theory of differential equations.}: if $p\in \overline \QQ \setminus
\overline \Gamma$ is a first singularity, then any compact set $X \subset %
\QQ \setminus \Gamma$ satisfying~\reef{DefnOfCompactlyGenerated} satisfies %
\beq  N \left(D^+(X) \setminus \{ p \} \right) = \infty \, . \eeq

\mysection{The $\D$-dimensional Bound}

\label{sec:Theorem}

Our equations of motion~\reef{GrEOM}, can be straightforwardly expressed in
terms of $\Omega, r$ and $\phi$. For convenience, we use the notation of
Dafermos~\cite{dafermos} adapted to $\D$ dimensions: \beqa   m &\equiv& 
\frac{r^{\D-3}}{2}\left(k+\frac{4\nu\lambda}{\Omega^2} \right) %
\labell{DefnOfM} \newline
\mu &\equiv& \frac{2}{r^{\D-3}} m = k +\frac{4\nu\lambda}{\Omega^2} %
\labell{DefnOfMu} \newline
\kappa &\equiv & \frac{\lambda}{k-\mu} \, . \labell{DefnOfKappa} \eeqa  With
these definitions, the equations of motion~\reef{GrEOM} can be written as %
\beqa   \lambda,_{u} = \nu_{,v} &=& \frac{2\lambda\nu}{k-\mu} \left[ (\D-3) 
\frac{m}{r^{\D-2}}  + \frac{4}{\D-2} r \frac{T_{uv}}{\Omega^2} \right] %
\labell{EOM12} \newline
m,_{u} &=& \frac{k-\mu}{\D-2} \left( \frac{r^{\D-2}T_{uu}}{\nu} -  2\frac{r^{%
\D-2}T_{uv}}{\lambda} \right) \labell{EOM3} \newline
m,_{v} &=& \frac{k-\mu}{\D-2} \left( \frac{r^{\D-2}T_{vv}}{\lambda}  -2\frac{%
r^{\D-2}T_{uv}}{\nu}\right) \labell{EOM4}
\eeqa  The equations of motion imply \beq   \kappa,_u = \frac{4}{\D-2}
\kappa \nu \frac{T_{uu}}{r} \labell{eq:kappa_u} \, . \eeq  Our aim is to
analyze these equations of motion along the lines of Ref.~\cite{dafermos} to
see how far (or under what additional assumptions) one can generalize
Theorem 3.1 of~\cite{dafermos}, which we now repeat here:

\begin{theorem}
Let $p\in \overline \QQ \setminus \QQ$ be a first singularity. Then either %
\beq  p \in \overline \Gamma \setminus \Gamma \eeq or \beq  J^-(p) \cap \QQ %
\cap D^+(X) \cap \TT \neq 0 \eeq for all compact $X$ satisfying~%
\reef{DefnOfCompactlyGenerated}. \label{Theorem}
\end{theorem}

Our strategy is to repeat the steps of the proof with the above definitions
and equations of motion and see how far we can progress.

Choose $\epsilon$ and $X$ as above in the definition of a first singularity.
First, we note that Eq.~\reef{eq:kappa_u} and the assumptions $\nu<0,
T_{uu}\geq 0$ imply \beq    \kappa,_u \leq 0\, ; \labell{eq:Kappa_uLt0} \eeq 
on the other hand, from the definition of $\kappa$ we have \beq    \kappa = -%
\frac{\Omega^2}{4 \nu} >0\, . \labell{eq:KappaGt0} \eeq  Since $\kappa$ is
positive, bounded on $X$, and cannot increase along the lines of constant $v$%
, it follows that is bounded on $D^+(X)\cap J^-(p)$ by a finite constant we
will denote by $K$, \beq    0 \leq \kappa \leq K\, . \labell{eq:KappaBnded} %
\eeq  Next, since $X$ is compact, the functions $r$, $\lambda$, $\nu$, $%
\lambda,_v, \lambda,_u=\nu,_v, \nu,_u, \Omega, \Omega,_u, \Omega,_v$ and $m$
are bounded on $X$. Let us denote the bounds as follows, \beqa    0 <r_0
\leq & r & \leq R\, , \labell{rBndedOnX} \newline
0 > \nu_0 \geq & \nu & \geq -N \, , \labell{NuBndedOnX} \newline
0 \leq &\lambda &< \Lambda \, , \labell{LambdaBndedOnX} \newline
& |m| &\leq M\, , \labell{mBndedOnX} \eeqa  \beq  |\lambda,_v|,
|\lambda,_u|, |\nu,_u|, |\Omega|, |\Omega^{-1}|, |\Omega,_u|, |\Omega,_v|
\leq H \, . \labell{DerivativesBndedOnX} \eeq (To simplify the notation, we
have used a common symbol $H$ to denote the upper bounds on the
derivatives.) The lower bounds on $r$ and $-\nu$ are a consequence of $\QQ$
being an open set and $X$ being its compact subset. We now obtain bounds on $%
r$, $\nu$, $\lambda$ and $m$ at any point with coordinates $u_\star, v_\star$
lying within $D^+(X)\cap J^-(p)$, analogous to ones Dafermos used in his
proof of the $\D=4$ theorem.

From~\reef{EOM3} we have (using $k-\mu >0, \nu<0$) \beqa    m,_u &\leq& -%
\frac{8}{(\D-2)} \frac{r^{\D-2}}{\Omega^2} T_{uv} (-\nu) \nn \newline
&\leq& \frac{8}{(\D-2)} r^{\D-2} (-\nu) C \, . \eeqa Integrating along $%
v=v_\star$ gives \beqa    m(u_\star, v_\star) & = & m(u_\epsilon, v_\star) +
\int_{u_\epsilon}^{u_\star} du\, m,_u \nn \newline
& \leq & M + \frac{8}{(\D-2)(\D-1)}CR^{\D-1} \nn \newline
& = & M + \tC \labell{mBndedAbove} \eeqa where we have introduced \beq \tC %
\equiv \frac{8}{(\D-2)(\D-1)}CR^{\D-1} \labell{DefnOftC} \eeq to simplify
notation in the following.

Similarly, from~\reef{EOM4} we infer \beqa    m _v &\geq& \frac{8}{(\D-2)} 
\frac{r^{\D-2}}{\Omega^2} T_{uv} \lambda \nn \newline
& \geq & - \frac{8}{(\D-2)} r^{\D-2} \lambda C \, . \eeqa Integrating along $%
u=u_\star$ we find \beqa    m(u_\star, v_\star) & = & m(u_\star, v_\epsilon)
+ \int_{v_\epsilon}^{v_\star} dv\, m,_v \nn \newline
& \geq & -M - \tC \, . \labell{mBndedBelow} \eeqa Thus $m$ is bounded both
from above and from below. It then follows that $\int du\, m,_u$ must be
bounded as well, namely it cannot be more than the difference between the
upper and the lower bound. Using~\reef{EOM3} again we obtain \beqa    
\lefteqn {2\left( M + \tC \right ) \geq \left| \int_{u_\epsilon}^{u_\star}
du\, m,_u \right|} \nn \newline
& = & \left| \int_{u_\epsilon}^{u_\star} du\, \frac{k-\mu}{\D-2} \left( 
\frac{r^{\D-2}T_{uu}}{\nu} -  2 \frac{r^{\D-2}T_{uv}}{\lambda} \right)
\right| \nn \newline
& = & \frac{2}{\D-2} \left|  \int_{u_\epsilon}^{u_\star} du\, (k-\mu)\frac{%
r^{\D-2}T_{uu}}{2\nu}  - \int_{T_{uv}>0} du\, (k-\mu)\frac{r^{\D-2}T_{uv}}{%
\lambda}  - \int_{T_{uv}<0} du\, (k-\mu)\frac{r^{\D-2} T_{uv}}{\lambda}
\right|\, . \nn \\ \mbox{} \labell{Inequality1} \eeqa  In the last line, the first and
second term are negative, while the last term is positive but bounded, \beq 
0 \leq - \frac{2}{\D-2} \int_{T_{uv}<0} du\,  (k-\mu)\frac{r^{\D-2} T_{uv}}{%
\lambda} \leq \tC \, . \labell{IntegralExpr0Bnded} \eeq  Thus we can write,
dropping the second term in expression~\reef{Inequality1}, \beq
\left|\int_{u_\epsilon}^{u_\star} du\, \frac{k-\mu}{\D-2} \frac{r^{\D%
-2}T_{uu}}{\nu} \right|  \leq  2M + 3\tC \, . \labell{IntegralExpr1Bnded} %
\eeq  Considering $\left| \int_{v_\epsilon}^{v_\star} dv\, m,_v \right|$,
one finds analogously \beq   \int_{v_\epsilon}^{v_\star} dv\, \frac{k-\mu}{\D%
-2} \frac{r^{\D-2}T_{vv}}{\lambda}  \leq  2M + 3 \tC \, . %
\labell{IntegralExpr2Bnded} \eeq  Next we turn to the quantity $\nu$.
Dividing~\reef{EOM12} by $\nu$ and integrating along a $u=u_\star$ curve we
find \beqa    | \nu(u_\star, v_\star) | & = & |\nu(u_\star, v_\epsilon) | 
\exp\left\{2\int dv\, \frac{\lambda}{k-\mu} \left[ (\D-3)\frac{m}{r^{\D-2}}
- \frac{4r}{\D-2} \frac{T_{uv}}{\Omega^2} \right] \right\} \nn \newline
&\leq& N \exp\left\{ 2K\epsilon \left[ \frac{\D-3}{r_0^{\D-2}} (M+\tC ) + 
\frac{4}{\D-2}CR \right] \right\} \nn \newline
& \equiv & N_b\, . \labell{NuBnded} \eeqa  For $\lambda$ we use~\reef{EOM12}
in its original form to obtain \beq  \lambda(u_\star, v_\star) \leq \Lambda
+ \left| 2\int du\, \kappa \nu \left[ (\D-3)\frac{m}{r^{\D-3}} - \frac{4}{\D%
-2} \frac{rT_{uv}}{\Omega^2} \right] \right| \, . \labell{LambdaBound} \eeq
Both $\kappa$ and $\nu$ are bounded as is the first term in the square
bracket. As for the second term in the square bracket, we first make a few
simple manipulations: \beq \frac{8}{\D-2} \left| \int du\,\kappa \nu \frac{%
rT_{uv}}{\Omega^2} \right|  \leq \frac{2}{\D-2} \frac{K}{r_0^{\D-3}} \left|
\int du\, (k-\mu)\frac{r^{\D-2} T_{uv}}{\lambda} \right| \, . %
\labell{Inequality3} \eeq For the last integral we use the inequality~%
\reef{Inequality1} again, this time together with~\reef{IntegralExpr1Bnded}.
We find \beq \left| \int du\, (k-\mu)\frac{r^{\D-2}T_{uv}}{\lambda} \right|
\leq (\D-2) \left(3M + 4\tC\right) \, . \labell{Inequality4} \eeq Putting it
all together, the final bound on $\lambda$ reads \beq  \lambda(u_\star,
v_\star) \leq \Lambda + 2KN \frac{\D-3}{D-2} \frac{M+\tC}{r_0^{\D-3}} + 2 
\frac{K}{r_0^{\D-3}} \left(3M + 4\tC\right) \, . \labell{LambdaBnded} \eeq
Lastly, from the relation $\Omega^2 = -4 \kappa \nu$ and the upper bounds~%
\reef{eq:KappaBnded} and~\reef{NuBnded} respectively for $\kappa$ and $\nu$
we obtain an upper bound for $|\Omega|$.

Thus, we have obtained bounds on the mass $m$, radius $r$ and its first
derivatives $\nu$ and $\lambda$, and $|\Omega|$ without having to consider
the matter sources in detail. These are not sufficient, however, to exclude
all singularities of the type Dafermos' work excluded -- indeed, we also
need to prove bounds on the second derivatives of $r$ (\ie first derivatives
of $\lambda$ and $\nu$) as well as on $|\Omega^{-1}|$ and the first
derivatives of $\Omega$. We now have two options on how to proceed further:
we either make additional assumptions on the geometry that will allow us to
continue with a general matter content, or we will specialize the matter
content and complete a proof of theorem~\ref{Theorem} by analyzing matter
evolution as well. The former route is followed in the next Section, while
the latter is discussed in Section~\ref{sec:Scalar}.

\mysection{Singularities in the Regular Region}

\subsection{General analysis}

\label{sec:GeneralSingularities}

We now make the additional assumption that the first singularity point $p$
of Theorem~\ref{Theorem} lies in the regular region $\RR$ so that $\lambda$
is bounded from below on $D^+(X)$: \beq  \lambda \geq \lambda_0 > 0 \mbox{ \
\ \ on \ } D^+(X) \, . \labell{AddtlLambdaBound} \eeq From the relation $%
\kappa = \lambda/(k-\mu)$ and the upper bound on $\kappa$ we infer that $%
k-\mu$ must be also bounded from below, \beq  k-\mu \geq \mu_0 >0 \, . %
\labell{KMuBndedBelow} \eeq From the definition~\reef{DefnOfMu} of $\mu$ and
the facts that $m$ is bounded from above and $r$ is bounded from below we
deduce that $k-mu$ is also bounded from above.

We now show that $|\nu|$ is also bounded from below by a positive number;
that will allow us to prove that $|\Omega|$ is bounded from below, implying
an upper bound on $|\Omega^{-1}|$.

To show the lower bound on $|\nu|$, we return to the integral in~%
\reef{NuBnded}. Obviously, to prove a lower bound on $|\nu|$, we must show
that the expression \beq  \int dv\, \frac{\lambda}{k-\mu} \left[ (\D-3)\frac{%
m}{r^{\D-2}} - \frac{4r}{\D-2} \frac{T_{uv}}{\Omega^2} \right] \eeq is
bounded. We have just shown that $\kappa = \lambda/(k-\mu)$ is bounded both
from above and from below; likewise, the first term in the square bracket in
the above expression is bounded. Therefore it is sufficient to show that the
integral \beq  \int dv\, \frac{T_{uv}}{\Omega^2} \labell{Integral5} \eeq is
bounded from above. To obtain this bound, we consider the analog of~%
\reef{Inequality1} written in terms of an integral over $v$: \beqa  
\lefteqn {2\left( M + \tC \right ) \geq \left| \int_{v_\epsilon}^{v_\star}
dv\, m,_v \right|} \nn \newline
& = & \left| \int_{v_\epsilon}^{v_\star} dv\, \frac{k-\mu}{\D-2} \left( 
\frac{r^{\D-2}T_{vv}}{\lambda} -  2 \frac{r^{\D-2}T_{uv}}{\nu} \right)
\right| \nn \newline
& = & \frac{2}{\D-2} \left|  \int_{v_\epsilon}^{v_\star} dv\, (k-\mu)\frac{%
r^{\D-2}T_{vv}}{2\lambda}  - \int_{T_{uv}>0} dv\, (k-\mu)\frac{r^{\D-2}T_{uv}%
}{\nu}  - \int_{T_{uv}<0} dv\, (k-\mu)\frac{r^{\D-2} T_{uv}}{\nu} \right|\,
. \nn \\ \mbox{} \labell{Inequality1v} \eeqa This time the first two terms in the last line
contribute with a positive sign while the last term contributes with a
negative sign. The (absolute value of the) last term is bounded (since $%
T_{uv}$ is bounded from below) analogously to~\reef{IntegralExpr0Bnded};
therefore the first and second term must be bounded as well. We are
interested in the latter; let us denote the bound as $I$: \beqa  I & > &
-\int_{T_{uv}>0} dv\, (k-\mu)\frac{r^{\D-2}T_{uv}}{\nu} \nn \newline
& \geq & - \int dv\, (k-\mu)\frac{r^{\D-2}T_{uv}}{\nu} \nn \newline
& = & \int dv\, 4\lambda r^{\D-2} \frac{T_{uv}}{\Omega^2}  \, . %
\labell{Inequality4v} \eeqa Since $r$ and $\lambda$ are bounded from below,
it follows that the expression~\reef{Integral5} is bounded, and thus $|\nu|$
is bounded from below.

The upper bound on $|\Omega^{-1}|$ now follows immediately from \beq 
\Omega^{-2} = -\frac{1}{4 \kappa \nu} \, , \labell{OmegaBnd} \eeq since both 
$\kappa$ and $\nu$ are bounded from below.

In summary, the additional assumption $\lambda \geq \lambda_0 >0$ allowed us
to prove that $|\nu|$ is bounded from below and $\Omega^{-1}$ is bounded. On
the other hand, deriving bounds on derivatives of $\Omega, \lambda$ and $\nu$
is still beyond reach. Indeed, the techniques we used only allow us to
estimate \emph{integrals} of various quantities that involve the stress
tensor; (integrable) divergences of local values of $T_{\mu\nu}$, and
therefore in the derivatives of $\Omega$, $\lambda$ and $\nu$ cannot be
excluded. Our result may then be summarized as forbidding the formation of 
\emph{non-integrable} (in the sense just described) naked singularities\footnote{%
An example of an integrable singularity is \eg a conical defect; on the
other hand, any singularity where the metric becomes degenerate (such as a Schwarzschild black hole
singularity) is non-integrable.} \emph{within the regular region}.

\subsection{Examples}

\label{sec:Examples}

In this subsection we investigate which matter theories satisfy the
assumptions~\reef{ConditionsOnT} on the matter stress--energy tensor.
Dafermos~\cite{dafermos} considered a scalar field only; in Section~\ref%
{sec:ScalarPlusGauge} we show that the assumptions~\reef{ConditionsOnT} are
satisfied by the theory of a scalar field coupled to a (non-abelian) gauge
field, and in Section~\ref{sec:Fluids} we look at matter described as a
perfect fluid.

\subsection{Interacting scalar and gauge fields}

\label{sec:ScalarPlusGauge}

In this subsection we show that the conditions~\reef{ConditionsOnT} are
satisfied in a general class of matter theories with the Lagrangian \beq \cL%
_{matter} = -\half \left|D \phi\right|^2 - V(\phi) - \frac{1}{4} \tr %
F_{\mu\nu}F^{\mu\nu} \, , \labell{GeneralMatterLagrangian} \eeq where the
gauge group is a unitary Lie group $G$, the scalar can be complex and
transforms in a representation $\cR$ of the group $G$, and the derivative $D$
is both diffeomorphism- and gauge-covariant. The scalar potential $V(\phi)$ is 
assumed to be continuous; the conditions~%
\reef{ConditionsOnT} will turn out to be satisfied if $V(\phi)$
is bounded from below. 

Before analyzing this extended system we note that scalar fields and 1-form
gauge fields are, up to Hodge duality, the most general that can respect $(\D%
-2)$--spherical symmetry in $\D$ dimensions. Higher-form potentials of rank
less than $\D-3$ have field strengths with too many components to fit into
the two dimensions spanned by $(u,v)$, while having too few components to
fill the $\D-2$ dimensions in a symmetric way. A $(\D-3)$--form potential
with a $(\D-2)$--form field strength can be spherically symmetric in $\D-2$
dimensions, but it is Hodge-dual to the Maxwell field strength $F_{\mu\nu}$,
while a $(\D-2)$--form potential with a $(\D-1)$--form field strength is
Hodge-dual to the scalar $\phi$ and its derivatives.

%The non-abelian field strength $F$ is given by
%\beq
%  F_{\mu\nu} = A_{[\mu},_{\nu]} - i [ A_\mu, A_\nu]
%\eeq
The stress tensor derived from the Lagrangian~\reef{GeneralMatterLagrangian}
is \beq  T_{\mu\nu} = -\half g_{\mu\nu} \left[ \half |D\phi|^2 + \frac{1}{4} %
\tr F^2 - V(\phi) \right]  + \half D_\mu \phi (D_\nu\phi)^\dagger + \half %
\tr F_{\mu\lambda}F_\nu{}^\lambda \, . \labell{GenStressT} \eeq Specializing
to components we find \beqa  T_{uu} &=& \half (D_u \phi) \cdot (D_u
\phi)^\dagger \, , \newline
T_{vv} &=& \half (D_v \phi) \cdot (D_v \phi)^\dagger \, , \newline
T_{uv} &=& -\half g_{uv} V(\phi) - \frac{1}{8} g^{uv} \tr F_{uv}F_{uv} \, . %
\eeqa In the $T_{uu}$ and $T_{vv}$ expressions, the dot denotes the
group-invariant scalar product within the representation $\cR$ in which $\phi
$ transforms under $G$. Assuming this inner product is positive (this
assumption is true for unitary groups), $T_{uu}$ and $T_{vv}$ are manifestly
positive. Likewise, given that $g_{uv}$ is negative, it is clear that the
first term in $T_{uv}$ will satisfy~\reef{ConditionsOnT} as long as the
potential $V$ satisfies $V(\phi) \geq -4C$, while the second term is
positive as long as the inner product in the adjoint representation of $G$
is positive. Thus the matter content specified by the Lagrangian~%
\reef{GeneralMatterLagrangian} satisfies the assumptions of our theorem. We
note that this statement also remains true if the scalar field develops a
vacuum expectation value and triggers a spontaneous gauge symmetry breaking.

\subsection{Perfect Fluids}

\label{sec:Fluids}

One can also contemplate matter in the case when only a hydrodynamic
description is available. Specifically, we can consider a perfect fluid with
a stress--energy tensor \beq  T_{\mu\nu} = (\rho+p)U_{\mu}U_{\nu} +
pg_{\mu\nu} \, , \labell{FluidStressT} \eeq where $U_\mu$ (not to be
confused with the coordinate $u$) is the velocity field of the fluid. We
will assume that the pressure $p$ and energy density $\rho$ are related 
\emph{via} an equation-of-state constant $w$ as \beq  p = w\rho\, . \eeq
Further, in accord with our general discussion in Section~\ref{sec:EOMs}, we
will assume that the velocity field $U^\mu$ has components only in the $(u,v)
$ coordinates. For a massive fluid, we then have \beqa  T_{uu} & = & (\rho +
p) U_u^2 \nn \newline
T_{vv} & = & (\rho + p) U_v^2 \labell{FluidTComponents}\newline
T_{uv} & = & -\half g_{uv} (\rho-p) \nn \eeqa where we have simplified the
expression for $T_{uv}$ using the fact that the velocity $U$ is time-like
and normalized to $U^2=-1$. It is clear that $T_{uu}$ and $T_{vv}$ will be
non-negative whenever either $(\rho \geq 0, w\geq -1)$ or $(\rho<0, w \leq
-1)$. On the other hand, $T_{uv}$ will remain non-negative as long as $(\rho
\geq 0, w\leq 1)$ or $(\rho\leq 0, w\leq -1)$. Together our theorem excludes
formation of naked singularities (non-integrable and in the regular region)
for any perfect fluid satisfying $\rho \geq 0, -1 \leq w \leq 1$. It would
be interesting to see whether a lower bound on $T_{uv}$ could be proved even
for (a range of) $w>1$ (with $\rho\geq 0$); however we were unable to do so.

For a massless fluid, the condition $0=U^2=2g_{uv}U^uU^v$ implies that at
least one of the components $U^u$, $U^v$ must be zero. $T_{uu}$ and $T_{vv}$
are still given by the expressions~\ref{FluidTComponents}, while $T_{uv}$
takes the form \beq  T_{uv} = pg_{uv} = -\half \Omega^2 p. \eeq Obviously, $%
T_{uv}$ will be bounded from below if the pressure $p$ is negative\footnote{%
Again, there may be a regime where even a positive pressure leads to a
development with a $T_{uv}$ bounded from below, but we were not able to find
such a development or conditions guaranteeing it.}. Taking into account the
condition $\rho+p \geq 0$ coming from the $T_{uu}$ and $T_{vv}$ components
we find that the assumptions of our theorem are satisfied when $(\rho>0, -1
\leq w \leq 0$).

\mysection{Specializing the Matter Content}

\label{sec:Scalar}

The result of Section~\ref{sec:GeneralSingularities} is that without a
detailed study of the evolution of matter that sources the geometry, only a
weaker statement about possible naked singularities can be made. On the
other hand, it is in general very hard to analyze a fully interacting
matter-gravity system such as the one studied in Section~\ref%
{sec:ScalarPlusGauge} to prove that its stress tensor will not develop
singularities of the kind that are not excluded our earlier general
arguments. Dafermos has nevertheless shown that such an analysis can be
performed for a scalar field in $\D=4$ dimensions; the aim of this Section
is to show that this analysis generalizes straightforwardly to any dimension 
$\D\geq 4$, any $k$ characterizing the topology of the transverse space and
that one can also include a free gauge field.

Thus, we analyze the evolution of a neutral scalar with potential $V$ and a
free gauge field potential $A_\mu$ with field strength $F_{\mu\nu}$ in more
detail, following Dafermos~\cite{dafermos}. The Lagrangian reads \beq \cL%
_{matter} = -\half \left(\nabla \phi\right)^2 - V(\phi) - \frac{1}{4} 
F_{\mu\nu}F^{\mu\nu} \, , \labell{MatterLagrangian} \eeq leading to the
stress--energy tensor \beqa  T_{\mu\nu} &=& -\frac{1}{\sqrt{g}} \frac{\delta
S_{matter}}{\delta g^{\mu\nu}} \nn \newline
& = & \half g_{\mu\nu} \left( -\half \left(\nabla \phi\right)^2 - V(\phi) - 
\frac{1}{4} F_{\mu\nu}F^{\mu\nu} \right)  + \half \phi,_\mu \phi,_\nu + %
\half F_{\mu\lambda} F_{\nu}{}^\lambda \, . \labell{MatterStressT} \eeqa For
the scalar field we again employ notation analogous to that used by Dafermos~%
\cite{dafermos}: we denote \beqa  \zeta &\equiv& r^\frac{\D-2}{2} \phi_{,u} %
\labell{ZetaDefn} \newline
\theta &\equiv& r^\frac{\D-2}{2} \phi_{,v} \, . \labell{ThetaDefn} \eeqa The
scalar equation of motion can then be written in two equivalent forms as %
\beqa  \zeta,_{v} &=& -\frac{\D-2}{2} \frac{\nu\theta}{r} - r^\frac{\D-2}{2} 
\frac{\lambda\nu}{k-\mu} V' \labell{EOM5} \newline
\theta,_{u} &=& -\frac{\D-2}{2} \frac{\lambda\zeta}{r} - r^\frac{\D-2}{2} 
\frac{\lambda\nu}{k-\mu} V' \, . \labell{EOM6} \eeqa As for the gauge field,
to respect spherical symmetry in any dimension $\D>4$, the field strength
can only have one nonzero component, namely $F_{uv}$, and it can only be a
function of $u$ and $v$. The gauge field equations, $F_{\mu\nu;}{}^\nu =0$,
can be written as \beq \left( \sqrt{g} F^{\mu \nu}\right),_{\nu} = 0\, . %
\labell{GaugeEOM2} \eeq In components they read \beq \left( \sqrt{g} F^{uv}
\right),_u = \left( \sqrt{g} F^{uv} \right),_v = 0\, , \eeq implying that %
\beq  F^{uv} = \frac{ q_e }{\Omega^2 r^{n-2}}\, . \labell{ElectricField} \eeq
The constant $q_e$ is, of course, interpreted as the electric charge.

In $\D=4$ dimensions one can have, in addition to the electric
configurations discussed above, also magnetic configurations: since the
internal space is 2-dimensional, one can take $F$ to be proportional to the
volume form on the internal space, $F_{ij} = q_m \sqrt{\gamma} \epsilon_{ij}$%
, where $q_m$ is the magnetic charge of the source. The most general
spherically symmetric field strength is a combination of the electric and
magnetic configurations with arbitrary charges.

Irrespective of the dimensionality it is clear that a spherically symmetric
gauge field has no propagating degrees of freedom. However, it will
contribute to the equations of motion \emph{via} the stress tensor $%
T_{\mu\nu}$. Both electric and magnetic charges contribute to the
stress--energy in the same way, making it useful to define a ``total''
(dyonic) charge \beq  q^2 \equiv q_e^2 + q_m^2\, . \eeq  This quantity will
enter the equations of motion we discuss next.

The matter stress tensor components in the frame~\reef{Metric} are \beqa 
T_{uu} & = & \half\frac{\zeta^2}{r^{\D-2}} \nn \newline
T_{vv} & = & \half\frac{\theta^2}{r^{\D-2}} \nn \newline
T_{uv} & = & \frac{1}{4} \Omega^2 \left( V(\phi) + \frac{1}{8} \frac{q^2}{%
r^{2(\D-2)}} \right) \, . \labell{StressTComponents} \eeqa The components $%
T_{uu}$ and $T_{vv}$ are manifestly non-negative, satisfying~%
\reef{ConditionsOnT}. The condition on $T_{uv}$ will be satisfied if \beq 
V(\phi) \geq -4C \, . \eeq In addition, we must require $V$ and its first
derivative to be continuous. Under these assumptions on $V(\phi)$ we can
derive bounds $\phi$ as well its derivatives (represented by $\theta$ and $%
\zeta$). Note that the presence or absence of the gauge field does not
change the condition on $V$ in a qualitative way: since $r$ is bounded from
both below and above, the contribution of the gauge field would at most lead
to a redefinition of the constant $C$.

Analogously to the discussion above~\reef{rBndedOnX}, compactness of $X$
together with continuity of $\phi$, $\zeta$ and $\theta$ implies that the
latter are bounded on $X$. We denote the bounds as follows, \beqa  |\phi|
&\leq & P \, ,\labell{PhiBndedOnX} \newline
|\theta| &\leq & \Theta \, , \labell{ThetaBndedOnX} \newline
|\zeta| &\leq & T \, ,\labell{ZetaBndedOnX} \eeqa Let us now consider $\phi$%
: \beqa    |\phi(u_\star, v_\star)| &\leq& |\phi(u_\star, v_\epsilon)|  +
\left| \int_{v_\epsilon}^{v_\star} dv\, \frac{\theta}{r^{(\D-2)/2}} \right| %
\nn \newline
&\leq& P + \left[ \int dv\, \frac{\theta^2(k-\mu)}{\lambda} \,  \int dv'\, 
\frac{\lambda}{r^{\D-2}(k-\mu)} \right]^{1/2} \, . \eeqa  The last
inequality follows from the well-known relation, for any two
square-integrable functions $f,g$, \beq \left( \int fg \right)^2 \leq
\left(\int f^2\right) \left(\int g^2 \right). \eeq The first term in the
square bracket above can be bounded using~\reef{IntegralExpr2Bnded}, while
the second can be bounded as \beq    \int dv\, \frac{\lambda}{r^{\D-2}(k-\mu)%
} \leq \epsilon K r_0^{2-\D}\, . \eeq  Together we have \beq  |\phi(u_\star,
v_\star)| \leq P + \left\{ (\D-2) ( 2M + 3\tC )  \epsilon K r_0^{2-\D}
\right\}^{1/2} \equiv P_b \, . \eeq  Having found a bound for $\phi$, we can
denote \beqa    C_b & \equiv & \sup_{|\phi|\leq P_b} V(\phi)\, , %
\labell{DefnOfC_b} \newline
C_b' & \equiv & \sup_{|\phi|\leq P_b} V'(\phi)\, . \labell{DefnOfC_bPrime} %
\eeqa  We can use the boundedness of the attained values of the potential to
constrain $\lambda$ more directly than in~\reef{LambdaBnded}. We use~%
\reef{EOM12} to obtain \beqa  | \lambda(u_\star, v_\star) | & \leq &
|\lambda(u_\epsilon, v_\star)| + \left| \int du\, \kappa \nu\left[ (\D-3)%
\frac{m}{r^{\D-2}} - \frac{2}{\D-2} rV \right] \right| \nn \newline
& \leq & \Lambda + 2K \left[ (\D-3) ( M + \tC) \frac{r_0^{3-\D}}{\D-3} + 
\frac{1}{\D-2} C_b R^2 \right]\, . \labell{LambdaBnded2} \eeqa Lastly, we
can derive bounds on the derivatives of $\phi$: integrating~\reef{EOM6}
along $v=v_\star$ we have \beq   |\theta(u_\star, v_\star)| \leq
|\theta(u_\epsilon, v_\star)| + \left|\frac{\D-2}{2} 
\int_{u_\epsilon}^{u_\star} du\, \frac{\lambda\zeta}{r} \right| +
\left|\int_{u_\epsilon}^{u_\star} du\, \frac{\lambda\nu}{k-\mu} r^{(\D-2)/2}
V' \right| \, . \labell{ThetaBound} \eeq  Eq.~\reef{ThetaBndedOnX} provides
a bound for the first term on the r.h.s, while for the second term we obtain %
\beqa  \left| \int_{u_\epsilon}^{u_\star} du\, \frac{\lambda\zeta}{r}
\right|  & = & \left| \int_{u_\epsilon}^{u_\star} du\, \left[ \sqrt{k-\mu} 
\frac{\zeta}{(-\nu)} \sqrt{-\nu} \right] \left[ \frac{\lambda}{\sqrt{k-\mu}} 
\frac{\sqrt{-\nu}}{r} \right] \right| \nn \newline
& \leq & \left[ \int_{u_\epsilon}^{u_\star} du\, (k-\mu)\left(\frac{\zeta}{%
\nu}\right)^2(-\nu) \right]^{1/2} \left[ \int_{u_\epsilon}^{u_\star} du'\, 
\frac{\lambda^2}{k-\mu} \frac{(-\nu)}{r^2} \right]^{1/2} \nn \newline
& = & \left[ \int_{u_\epsilon}^{u_\star} du\, \left(-2(\D-2) m,_u +16 \nu 
\frac{r^{\D-2}T_{uv}}{\Omega^2} \right) \right]^{1/2} \left[
\int_{u_\epsilon}^{u_\star} du'\, \kappa^2 (k-\mu) \frac{(-\nu)}{r^2} \right]%
^{1/2} \nn \newline
&\leq & \left[ 4(\D-2) ( M + \tC ) + \frac{16}{\D-1}CR^{\D-1} \right]^{1/2} %
\nn \newline
& & \times \left[ K^2 \left\{ |k| + \frac{1}{r_0^{\D-3}} ( M + \tC )
\right\} \frac{1}{r_0} \right]^{1/2} \eeqa   (we have used~\reef{EOM3} and
the lower bound on $T_{uv}$ in~\reef{ConditionsOnT}), and for the third term
we have \beq \left|\int_{u_\epsilon}^{u_\star} du\, \frac{\lambda\nu}{k-\mu}
r^{(\D-2)/2} V' \right| \leq \frac{2}{\D} KC_b'R^{\D/2}\, . \eeq Since all
three terms on the r.h.s. of~\reef{ThetaBound} are bounded, we can write %
\beq  |\theta(u_\star, v_\star)| \leq \Theta_b\, . \labell{ThetaBnded} \eeq 
For $\zeta$ we have, from~\reef{EOM5}, \beqa  | \zeta(u_\star, v_\star) |
&\leq& |\zeta(u_\star, v_\epsilon) | + \left| \int dv\, \left( \frac{\D-2}{2r%
} \nu\theta + r^{\D-2}{2} \frac{\lambda \nu}{k-\mu} V' \right) \right| \nn 
\newline
& \leq & Z_b \, . \labell{ZetaBnded} \eeqa The bound can by now be arrived
at in different ways, the simplest one being to notice that all quantities
entering the r.h.s of the first line in~\reef{ZetaBnded} are bounded ($r$
entering denominators is bounded from below and $\lambda/(k-\mu) \equiv
\kappa < K$, so denominators do not pose any problems either).

We have thus established that $\phi$ as well as its first derivatives are
finite. It follows (from continuity of $V(\phi)$) that all scalar
contributions to the stress--energy tensor are bounded; the gauge
contribution is bounded by virtue of the lower bound on $r$.

Having thus established finiteness of the matter stress--energy tensor, it
is now clear from~\reef{NuBnded} that $|\nu|$ will be bounded from below as
well as from above. Further, integrating Eq.~\reef{eq:kappa_u} as an
equation for $\kappa$ with respect to $u$ similarly shows that $\kappa$ will
be bounded from below as well. Eq.~\reef{OmegaBnd} then gives an upper bound
on both $|\Omega|$ and $|\Omega^{-1}|$.

To establish a bound on $\nu,_u$ we differentiate~\reef{EOM12} with respect
to $u$ to obtain \beqa  (\nu,_u),_v &=& 2(\nu,_u \kappa + \nu \kappa,_u) %
\left[ (\D-3) \frac{m}{r^{\D-2}}  + \frac{1}{\D-2} r \left( V(\phi) + \frac{1%
}{8} \frac{q^2}{r^{2(\D-2)}} \right) \right] \nn \newline
& & 2\nu \kappa \left[ (\D-3) \frac{m,_u}{r^{\D-2}} - (\D-2)(\D-3)\nu \frac{m%
}{r^{D-1}}  + \frac{1}{\D-2} \nu \left( V(\phi) + \frac{1}{8} \frac{q^2}{%
r^{2(\D-2)}} \right) \right. \nn \newline
& & \left. + \frac{1}{\D-2} r \left(V'(\phi) r^{\frac{2-n}{2}}\zeta  + \frac{%
2(\D-2)}{8} \nu \frac{q^2}{r^{2\D-3)}} \right) \right] \eeqa Viewing this
equation as a differential equation for $\nu,_u$ (taking into account~%
\reef{EOM3} and~\reef{eq:kappa_u} with the stress--energy tensor given by~%
\reef{StressTComponents}), it is easy to see that all coefficients are
uniformly bounded. Thus, integrating this equation in $v$ with regular
initial conditions must give a uniformly bounded $\nu,_u$. A bound on $%
\lambda,_v$ is derived analogously.

To derive bounds on the derivatives of $\Omega$, we use the internal ($ij$)
components of~\ref{GrEOM} that lead to \beq  -(\ln \Omega^2),_{uv} = - \frac{%
(\D-3)(4-\D)}{r^2}\left[ r,_u r,_v k\frac{\Omega^2}{2}\right] + 2 \frac{%
\zeta\theta}{r^{\D-2}} - V(\phi) + \frac{q^2}{8r^{2(n-2)}} \, . %
\labell{OmegaEOM} \eeq Integrating this equation in $u$ and $v$ will give a
bound on $\Omega,_v$ and $\Omega,_u$, respectively, since all terms on the
r.h.s. are bounded (and initial values of $\Omega,_v$ and $\Omega,_u$ on $X$
are bounded by assumption~\reef{DerivativesBndedOnX}.

This completes the analog of the $\D=4$ proof given by Dafermos for all
dimensions 4 and greater, and for locally flat and locally hyperbolic
transverse part of the spacetime.

\mysection{Comments on certain problems involving potentials unbounded from
below}

\label{sec:comments}

Unfortunately the theorem proved here does not apply to an interesting class
of problems, namely the 5-dimensional problems studied in~\cite{HHM2, HH2}
that contain a scalar field with a potential unbounded from below. It would
be surprising if our methods extended to that case: while Dafermos' theorem
is in its nature local and thus independent of the boundary conditions
imposed on the scalar field, the superstring-inspired supergravity theories
of~\cite{HHM2, HH2} depend crucially on the scalar boundary conditions.
Indeed, they would be ill-defined if the scalar field were not required to
approach asymptotically a fixed value for which the potential has a maximum.
It follows that a proof along the lines of the one discussed here would then
also apply to unstable gravitational theories -- but in such theories one
would expect naked singularities to form, thus invalidating the putative
proof of the theorem.

In the absence of a proof for theories with unbounded potentials, one can
ask whether the Dafermos theorem can be used at least as a qualitative
guide; in other words, whether qualitative features of the evolution of
theories with potentials unbounded from below actually depend on the fact
that the potential is unbounded. One can, for example, imagine cutting the
potential off at some large but finite (negative) value and ask whether this
change affects the basic properties of the solutions discussed in~\cite%
{HHM2, HH2}.

The solutions discussed in~\cite{HHM2, HH2} contain a homogeneous inner
region that evolves as (a patch of) an FRW universe. As pointed out in~\cite%
{HHM2}, the evolution near the singularity is dominated by the kinetic
energy of the scalar field. In particular, cutting off the potential at a
large negative value does not qualitatively change the evolution; a
space-like singularity still forms in finite proper time. If the potential
is cut off from below\footnote{%
The cutoff must be at least continuously differentiable to satisfy the
assumptions of the theorem; one can certainly use a smooth cutoff if
necessary.}, the theorem proved above forbids the type of singularity
suggested in~\cite{HHM2}.

Unfortunately, this argument does not prove that a naked singularity cannot
form in the setup of~\cite{HHM2}: our argument just shows that if a
singularity forms, in the inner region it will appear as a big crunch-type
singularity. However the argument does not apply to the intermediate region,
where the singularity could become time-like or end. Only if one could show
that the potential cutoff is inconsequential even in this region, would the
singularity of \cite{HHM2} be ruled out and the conclusion of~\cite{HH2}
(that such singularities are of the big crunch type, cutting off all space)
would be confirmed.

The situation is better in the case of the gravitational instanton of~\cite%
{HH2}. As discussed in that work, there is no ``intermediate region'' --
namely, outside of the lightcone of the origin (for details, see~\cite{HH2}
Sec. 3.3 and 3.4) the scalar field remains bounded and as such will be
unaffected if the potential is cut off only for sufficiently large $\phi$.
On the other hand, the spacetime inside the lightcone is an open FRW
universe that will evolve to a big crunch irrespective of whether the
potential is bounded from below. In this case, therefore, our generalization
of Dafermos' theorem confirms that a naked singularity cannot form.

\mysection{Acknowledgements}

The authors would like to thank David Garfinkle and Mihalis Dafermos for
insightful discussions and encouragement. This research was supported by in
part by the Natural Sciences and Engineering Research Council of Canada.

\appendix

\end{document}